\documentclass[12pt]{article}
\usepackage{epsfig,graphicx}
\usepackage{fpcp03}

\def\barD{\overline D{}^0}

\def\Bbar{\overline{B}}
\def\DDbar{D^0-\overline D{}^0}
\def\BBbar{B^0-\overline B{}^0}

\def\D0bar{\overline D{}^0}
\def\K0bar{\overline K{}^0}

\def\cal{{\it}}

\def\beq{\begin{equation}}
\def\eeq{\end{equation}}
\def\beqa{\begin{eqnarray}}
\def\eeqa{\end{eqnarray}}
\def\bea{\begin{eqnarray}}
\def\eea{\end{eqnarray}}

\def\beq{\begin{equation}}
\def\eeq#1{\label{#1}\end{equation}}
\def\eeqn{\end{equation}}

\def\beqa{\begin{eqnarray}}
\def\eeqa#1{\label{#1}\end{eqnarray}}
\def\eeqan{\end{eqnarray}}

\let\bar=\overbar

\def\Dslash{\not{\hbox{\kern-4pt $D$}}}
\def\dslash{\not{\hbox{\kern-2pt $\del$}}}

\def\msb{{\bar{\ssstyle M \kern -1pt S}}}

\def\BB0bar{B^0 {\overline B}^0}
\def\BB0dbar{B_d^0 {\overline B}_d^0}
\def\BB0sbar{B_s^0 {\overline B}_s^0}

\RequirePackage{xspace}

\usepackage{relsize}
\def\babar{\mbox{\slshape B\kern-0.1em{\smaller A}\kern-0.1em
    B\kern-0.1em{\smaller A\kern-0.2em R}}}

\def\Kbar  {\kern 0.2em\overline{\kern -0.2em K}{}\xspace}

\def\Kz    {\ensuremath{K^0}\xspace}
\def\Kzb   {\ensuremath{\Kbar^0}\xspace}
\def\KzKzb {\ensuremath{\Kz \kern -0.16em \Kzb}\xspace}
\def\Kp    {\ensuremath{K^+}\xspace}
\def\Km    {\ensuremath{K^-}\xspace}

\def\KpKm  {\ensuremath{\Kp \kern -0.16em \Km}\xspace}

\def\Dbar    {\kern 0.2em\overline{\kern -0.2em D}{}\xspace}

\def\Dz      {\ensuremath{D^0}\xspace}
\def\Dzb     {\ensuremath{\Dbar^0}\xspace}
\def\DzDzb   {\ensuremath{\Dz {\kern -0.16em \Dzb}}\xspace}
\def\Dp      {\ensuremath{D^+}\xspace}
\def\Dm      {\ensuremath{D^-}\xspace}

\def\DpDm    {\ensuremath{\Dp {\kern -0.16em \Dm}}\xspace}

\def\Bbar    {\kern 0.18em\overline{\kern -0.18em B}{}\xspace}

\def\BB      {\ensuremath{B\Bbar}\xspace} 
\def\Bz      {\ensuremath{B^0}\xspace}
\def\Bzb     {\ensuremath{\Bbar^0}\xspace}
\def\BzBzb   {\ensuremath{\Bz {\kern -0.16em \Bzb}}\xspace}
\def\Bu      {\ensuremath{B^+}\xspace}
\def\Bub     {\ensuremath{B^-}\xspace}

\def\BpBm    {\ensuremath{\Bu {\kern -0.16em \Bub}}\xspace}

\mathchardef\Upsilon="7107
\def\Y#1S{\ensuremath{\Upsilon{(#1S)}}\xspace}

\mathchardef\Deltares="7101
\mathchardef\Xi="7104
\mathchardef\Lambda="7103
\mathchardef\Sigma="7106
\mathchardef\Omega="710A

\def\Deltabar{\kern 0.25em\overline{\kern -0.25em \Deltares}{}\xspace}
\def\Lbar{\kern 0.2em\overline{\kern -0.2em\Lambda\kern 0.05em}\kern-0.05em{}\xspace}
\def\Sigbar{\kern 0.2em\overline{\kern -0.2em \Sigma}{}\xspace}
\def\Xibar{\kern 0.2em\overline{\kern -0.2em \Xi}{}\xspace}
\def\Obar{\kern 0.2em\overline{\kern -0.2em \Omega}{}\xspace}
\def\Nbar{\kern 0.2em\overline{\kern -0.2em N}{}\xspace}
\def\Xb{\kern 0.2em\overline{\kern -0.2em X}{}\xspace}

\newcommand{\tev}{\ensuremath{\mathrm{\,Te\kern -0.1em V}}\xspace}
\newcommand{\gev}{\ensuremath{\mathrm{\,Ge\kern -0.1em V}}\xspace}
\newcommand{\mev}{\ensuremath{\mathrm{\,Me\kern -0.1em V}}\xspace}
\newcommand{\kev}{\ensuremath{\mathrm{\,ke\kern -0.1em V}}\xspace}
\newcommand{\ev}{\ensuremath{\mathrm{\,e\kern -0.1em V}}\xspace}
\newcommand{\gevc}{\ensuremath{{\mathrm{\,Ge\kern -0.1em V\!/}c}}\xspace}
\newcommand{\mevc}{\ensuremath{{\mathrm{\,Me\kern -0.1em V\!/}c}}\xspace}
\newcommand{\gevcc}{\ensuremath{{\mathrm{\,Ge\kern -0.1em V\!/}c^2}}\xspace}
\newcommand{\mevcc}{\ensuremath{{\mathrm{\,Me\kern -0.1em V\!/}c^2}}\xspace}

\def\mus  {\ensuremath{\rm \,\mus}\xspace}

\def\mus        {\ensuremath{\,\mu{\rm s}}\xspace}    

\def\to                 {\ensuremath{\rightarrow}\xspace}

\def\pep2{PEP-II}

\def\gsim{{~\raise.15em\hbox{$>$}\kern-.85em
          \lower.35em\hbox{$\sim$}~}\xspace}
\def\lsim{{~\raise.15em\hbox{$<$}\kern-.85em
          \lower.35em\hbox{$\sim$}~}\xspace}

\xspace

\def\jetset74   {\mbox{\tt Jetset \hspace{-0.5em}7.\hspace{-0.2em}4}\xspace}

\begin{document}


\Title{Charm physics: theoretical review}
\bigskip


%
\label{PetrovStart}

%
\author{ Alexey A. Petrov\index{Petrov, A.~A.} }

%
\address{Department of Physics and Astronomy\\
Wayne State University\\
Detroit, MI 48201 USA\\
}

\makeauthor\abstracts{
We review recent developments in charm physics, focusing on the physics of charmed 
mesons. We discuss charm spectroscopy, decay constants, as 
well as searches for new physics with charmed mesons. We discuss $\DDbar$ mixing 
and CP-violation in charm decays. We also present the modified Nelson plot of charm 
mixing predictions.
}

\section{Introduction}

Charm physics plays a unique dual role in the modern investigations of flavor physics. 
Charm decay and production experiments provide valuable checks and supporting measurements 
for studies of CP-violation in measurements of CKM parameters in b-physics, as 
well as outstanding opportunities for searches for new physics.
Historically, many methods of heavy quark physics have been first tested in charmed hadrons.
The fact that a b-quark mainly decays into a charm quark makes charm physics an integral
part of any b-physics program. In many cases, direct measurements of charm decay parameters 
directly affect the studies of fundamental electroweak physics in B decays~\cite{Petrov:2003rm}.

This year brought several interesting developments in some seemingly well-understood 
sectors of charm physics, such as meson spectroscopy. Here I shall discuss theoretical
implications of these and other results. The experimental status of charmed meson 
spectroscopy was discussed in R.~Chistov's talk~\cite{chistov-fpcp03}. Resent results 
in the measurements of charmed meson formfactors, lifetimes and $\DDbar$ mixing
parameters were discussed by W.~Johns~\cite{johns-fpcp03} and G.~Boca~\cite{boca-fpcp03}.

\section{Spectroscopy}

Meson spectroscopy is an important laboratory for understanding quark 
confinement. Mesons containing one heavy quark can provide 
valuable information about the structure of the QCD Lagrangian,
as spectroscopic considerations simplify significantly in the 
limit of infinitely heavy quark, $m_Q/\Lambda \to \infty$, where 
$\Lambda$ represents a typical scale of hadronic interactions.
While charm quark hardly satisfies this conditions, it is nethertheless
useful to apply these considerations to the charmed quark systems.
In this limit the heavy quark spin $S_Q$ decouples, so the total 
angular momentum of the light degrees of freedom 
$J_l^p$ becomes a ``good'' quantum number. Since parity of a meson can be 
obtained by knowing the angular momentum quantum number $l$ as $(-1)^{l+1}$,
this leads to an important prediction of heavy quark symmetry: the appearance 
of heavy meson states in the form of degenerate parity doublets classified by the 
total angular momentum of the light degrees of freedom (see Table~(\ref{tab1})),
\begin{equation}
S^p=J_l^p\pm \frac{1}{2}.
\end{equation}
This mass degeneracy is lifted with the inclusion of subleading $1/m_Q$ 
corrections. This useful picture is built into many quark-model descriptions 
of heavy meson spectra. The resulting models have been very successful in 
explaining the spectrum 
of negative-parity scalar and vector $J_l^p=1/2^-$ and positive-parity 
vector and tensor $J_l^p=3/2^+$ states.
\begin{table}[htbp]
\begin{center}
\begin{tabular}{|c|c|c|c|c|c|}
\hline
$L$ & 0 & \multicolumn{2}{|c|}{1} & \multicolumn{2}{|c|}{2}\\ 
\hline
$J_l$ & $1/2$ & $1/2$ & $3/2$ & $3/2$ & $5/2$ \\
$S$   & $0,1$ & $0,1$ & $1,2$ & $1,2$ & $2,3$ \\
\hline
\end{tabular}
\caption{Total angular momentum assignments for heavy-light mesons} 
\label{tab1}
\end{center}
\end{table}
A narrow resonance in $D_s^+\pi^0$ was recently reported by BaBar~\cite{dspi0_babar} 
and confirmed by the 
CLEO~\cite{dspi0_cleo} and Belle~\cite{dspi0_belle} collaborations. Its decay 
patterns suggest a quark-model $0^+$ classification, which would 
identify it with the positive-parity $J_l^p=1/2^+$ $p$-wave state. As in the 
$D$ meson system, $p$-wave states for the $D_s^+$ system are expected, and two 
narrow states, $D_{s1}(2536)$ and $D_{s2}(2573)$ were discovered by ARGUS and 
CLEO, respectively~\cite{PDG}. 
In analogy to the $D$ system, two broad states are also expected.

The mass of the new state $2317.6\pm 1.3$ MeV appears surprisingly low and its width appears 
to be too small for quark model practitioners. In fact, this state appears below 
$DK$ threshold, closing off the most natural decay channel for this state. 
This forces it to decay mainly via an
isospin-violating transition into the $D_s^+\pi^0$ final state which makes its width quite 
narrow. Its mass disagrees with most predictions of quark 
models~\cite{godfrey_isgur,eichten,Gupta:1994mw,Zeng:1994vj,Kalashnikova:2001ig,Ebert:1997nk}
available prior to its observation.
For example, a mass of $2487$ MeV is obtained in the potential model 
calculation by Eichten and Di Pierro~\cite{eichten}. Quenched lattice calculations 
also seem to favor larger values of the mass of this state~\cite{Lewis:2000sv} 
(see, however, \cite{lattice}). 
\begin{table}[htbp]
\begin{center}
\begin{tabular}{|l|c|c|}
\hline
Reference & $0^+$ mass & $1^+$ mass \\ 
\hline
Ebert et al (98)~\cite{Ebert:1997nk} & $2.51$~GeV & $2.57$~GeV\\
Godfrey-Isgur (85)~\cite{godfrey_isgur} & $2.48$~GeV & $2.55$~GeV\\
DiPierro-Eichten (01)~\cite{eichten} & $2.49$~GeV & $2.54$~GeV\\
Gupta-Johnson (95)~\cite{Gupta:1994mw} & $2.38$~GeV & $2.52$~GeV\\
Zeng et al (95)~\cite{Zeng:1994vj} & $2.38$~GeV & $2.51$~GeV\\
Experiment & $2.317$~GeV & $2.463$~GeV\\
\hline
\end{tabular}
\caption{Theoretical predictions for masses of $0^+$ and $1^+$ $D_s$ states} 
\label{tab2}
\end{center}
\end{table}
This led to a lively discussion of the possible non-$q\bar q$ nature
of this state~\cite{barnes_lipkin,cheng_hou,szcz,cahn_jackson}.
A possibility of a state that is an admixture of a four-quark state and 
a $q\bar q$ states was discussed in~\cite{Browder:2003fk}.
In addition, a second narrow state is observed in $D_s^{*+}\pi^0$ at a mass near
2460 MeV~\cite{dspi0_cleo,dspi0_belle}. 
This state would naturally be identified as a spin 1 positive parity 
$p$-wave meson. However, its mass also appears too low for the potential model 
expectations (e.g. 2605 MeV~\cite{eichten,Lucha:2003gs}, see also Table~\ref{tab2}).
Its radiative decays to the ground state $D_s$ meson were observed with
\begin{equation}
\frac{\Gamma(D_{sJ}(2460) \to D_s \gamma)}{\Gamma(D_{sJ}(2460) \to D_s \pi^0)}=
0.44\pm 0.10,
\end{equation}
while doubly charged states were not observed in $D_s^{\pm}\pi^\pm$ channels.
Finally, $D_{sJ}$ states were also observed in B-decays $B \to D D^{(*)}_{sJ}$.

The low values of the masses for these states, however, do
not signal a breakdown of quark-model descriptions of the heavy meson 
spectrum, as it is difficult to assess the accuracy of these
predictions, especially in the charm sector. Many authors make use of the 
non-relativistic nature of the charm quark, taking into account
$1/m_c$ corrections only. For the $0^+$ state, quark model predictions 
range from the values of $2387-2395$~MeV~\cite{Gupta:1994mw,Kalashnikova:2001ig} 
(still above the $DK$ threshold) on the low end of the spectrum to 
$2508$~MeV~\cite{Ebert:1997nk} on the high end. Since the described 
phenomena are highly non-perturbative, one should be careful before 
making a judgment on the nature of a given state based solely on the 
prediction of a given quark model. For example, as discussed above, in 
the $m_c \to \infty$ limit the $0^+$ and $1^+$ states are expected to 
become degenerate in mass, $m_{0^+},m_{1^+} \to M$. This can be
emulated in quark models by neglecting heavy-quark symmetry-violating $1/m_c$ 
corrections. Yet, different quark models predict very different
behavior in this "heavy-quark limit": for instance, one potential 
model~\cite{Gupta:1994mw} predicts that the mass $M$ of the 
$(0^+, 1^+)$ multiplet will decrease to approximately $2382$~MeV
(which is less than the mass of the $0^+$ state predicted in this model with
the full potential), while in a QCD string
model~\cite{Kalashnikova:2001ig} 
it is expected to increase up to $2500$~MeV (which is much greater
than the mass of the $0^+$ state predicted in this model with the full 
potential). In addition, quark models, modified to include 
chiral symmetry constraints, generally predicted lower values of mass 
splitting between $(0^-,1^-)$ and $(0^+,1^+)$ multiplets, of the order
of $200-300$~MeV~\cite{chiral}. In addition, one has to remember that most of 
the unusial details about these states, such as narrowness of their decay widths,
simply follows from the fact that the mass of that state is smaller than 
the $D^{(*)}K$ threshold. It is then only the fact that the new state 
appears below $D K$ threshold and is almost degenerate with a
non-strange $0^+$ $p$-wave $D$ state~\cite{ddst_belle} is curious and deserves an 
investigation, although could be purely accidental.

A combination of experimental measurements described above can shed some light onto 
the nature of these states. For instance, molecular-type explanation of the low masses 
of these states implies the existence of the doubly-charged states, which were not 
observed. On the contrary, a possible disagreement of the observed branching ratios 
of $B$ decays into these states with calculations of their branching ratios in naive 
factorization could favor molecular nature of these states~\cite{Datta:2003re}. But
it could as well signal a breakdown of naive factorization in $B$ decays into the 
pair of $0^-,0^+(1^+)$ open-charm states, which was never really tested, or simply
reflect our ignorance of the decay constants of positive parity mesons~\cite{veseli}.
The observation of the radiative decay of $D_{sJ}(2460)$ favors $q\bar q$ (or maybe
$q\bar q$-four-quark-state admixture) explanation of the nature of these mesons.

\section{Decay constants and B-physics experiments.}

Since $m_b,~m_c \gg \Lambda_{QCD}$, both charm and bottom quarks can be regarded as
heavy quarks~\footnote{This approximation obviously works better for bottom than for 
charm quarks.}.
Naturally, heavy quark symmetry relates observables in B and D transitions.
As an example, let us consider measurements in the charm sector
affect determinations of the CKM matrix elements relevant to top quark in
$\BBbar$ mixing.

A mass difference of mass eigenstates in $\BBbar$ system can be written as
\begin{equation}
\label{offdiag}
\Delta m_d =
\!\! {\cal C}\left[\alpha_s^{(5)}(\mu)\right]^{-6/23}
\left[1+\frac{\alpha_s^{(5)}(\mu)}{4 \pi} J_5 \right]
\langle\bar B_d^0|{\cal O}(\mu)|B_d^0\rangle, \nonumber
\end{equation}
where ${\cal C}=G_F^2M_W^2
\left({V_{tb}}^{*}V_{td}\right)^2 \eta_B m_B S_0(x_t)/\left(4 \pi^2\right)$ (see
Ref.~\cite{Buchalla:1995vs} for complete definitions of the parameters in this expression
).
The largest uncertainty of about 30\% in the theoretical calculation
is introduced by the poorly known hadronic matrix element
$
{\cal A} = \langle\bar B^0|{\cal O}(\mu)|B^0\rangle
$.
Evaluation of this matrix element is a genuine non-perturbative task, which can
be approached with several different techniques. The simplest approach
(``factorization'')
reduces the
matrix element ${\cal A}$ to the product of matrix elements measured in
leptonic $B$ decays
$
{\cal A}^{f} = (8/3)
$
$
\langle\bar B^0|\bar b_L\gamma_{\sigma}d_L|0\rangle
\langle 0|\bar b_L \gamma^{\sigma}d_L|B^0\rangle = (2/3) f_B^2 m_B^2
$, where we employed the definition of the decay constant $f_B$,
\begin{equation}
\langle 0|\bar b_L \gamma_\mu d_L|B^0({\bf p})\rangle = i p_\mu f_{B}/2.
\end{equation}
A deviation from the factorization ansatz is usually described by the parameter
$B_{B_d}$ defined as
$
{\cal A} = B_{B_d} {\cal A}^{f}
$;
in factorization $B_{B_d}=1$. Similar considerations lead to an introduction of
the parameter $B_{B_s}$ defined for mixing of $B_s$ mesons. It is important to
note that the parameters $B_{B_q}$ depend on the chosen renormalization scale
and scheme. It is convenient to introduce renormalization-group invariant
parameters $\hat B_{B_q}$
\begin{equation}
\label{Bhat}
\hat B_{B_q} =
\!\! \left[\alpha_s^{(5)}(\mu)\right]^{-6/23}
\left[1+\frac{\alpha_s^{(5)}(\mu)}{4 \pi} J_5 \right]
B_{B_q}.
\end{equation}
We provide averages of $\hat B_{B_q}$, as well as the ratio
$\hat B_{B_s}/\hat B_{B_d}$ from the review~\cite{Battaglia:2003in} as well
as from two more recent evaluations~\cite{Aoki:2003xb,Korner:2003zk} in
Table~\ref{table1}.
Thus, at least naively, one can determine CKM matrix element $V_{td}$ by measuring $f_B$
and $\Delta m_d$ and computing $B_{B_q}$.

This direct approach, however, meets several difficulties. First,
leptonic decay constant $f_B$ can in principle be extracted from leptonic decays of
charged B mesons. The corresponding decay width is
\begin{equation}
\Gamma(B \to l \nu) = \frac{G_F^2}{8\pi} f_B^2 \left|V_{ub}\right|^2 m_l^2 m_B
\left(1-\frac{m_l^2}{m_B^2}\right).
\end{equation}
This width is seen to be quite small due to the smallness of the CKM factor
$\left|V_{ub}\right|$ and helicity suppression factor of $m_l^2$. In addition,
experimental difficulties are also expected due to the backgrounds stemming from
the presence of a neutrino in the final state.

\begin{table}[htbp]
\begin{center}
\begin{tabular}{|c|c|c|}
\hline
Method (reference) & $\hat B_{B_d}$ & $\hat B_{B_s}/\hat B_{B_d}$ \\
\hline\hline
Lattice, '03~\cite{Battaglia:2003in} & $1.34(12)$ &
$1.00(3)$ \\
QCDSR, '03~\cite{Battaglia:2003in} & $1.67\pm 0.23$ &
$\approx 1$ \\
Lattice, '03~\cite{Aoki:2003xb} & $1.277(88)(^{+86}_{-95})$ &
$1.017(16)(^{+56}_{-17})$ \\
QCDSR, '03~\cite{Korner:2003zk} & $1.60\pm 0.03$ &
$\approx 1$ \\
\hline
\end{tabular}
\caption{\label{table1} Renormalization-group independent B-parameters.}
\end{center}
\end{table}

Second, computation of $B_{B_q}$ is quite difficult and requires the use of non-perturbative
techniques such as lattice or QCD Sum Rules. Current uncertainties in the determinations
of $f_B$ and $B_{B_q}$ are quite large. It turns out that evaluation of
the ratio of
\begin{equation}
\frac{\Delta m_d}{\Delta m_s} = \frac{m_{B_d}}{m_{B_s}}
\left[\frac{\sqrt{B_{B_d}}f_{B_d}}{\sqrt{B_{B_s}}f_{B_s}}\right]^2
\left|\frac{V_{td}}{V_{ts}}\right|^2,
\end{equation}
is favored by lattice community, as many systematic errors cancel in this ratio.
This gives a ratio of $\left|V_{td}/V_{ts}\right|$, which provides a non-trivial
constraint on CKM parameters in the $\rho-\eta$ plane.

Instead, one can make use of ample statistics available in charm production
experiments, as heavy quark and $SU(3)$ flavor symmetries relate the ratio
of charm decay constants $f_{D_s}/f_{D}$ to beauty decay constants
$f_{B_s}/f_{B}$
\begin{equation}
\frac{f_{B_s}/f_{B}}{f_{D_s}/f_{D}}=1+{\cal O}(m_s)\times{\cal O}(1/m_b-1/m_c).
\end{equation}
Note that SU(3)-violating corrections can also be evaluated in chiral perturbation
theory~\cite{Grinstein:1993ys}. One still needs to rely on the theoretical
determination of $B_{B_q}$.

Similar techniques of relating $B$ and $D$ decays can also be used to extract other
CKM matrix elements, like $V_{ub}$~\cite{Ligeti:1997aq}, studies of lifetime
patterns of heavy hadrons~\cite{Pedrini:2003ee}, and tuning lattice QCD
calculations~\cite{Juttner:2003cf}.

\section{Charm mixing and CP violation}

One of the important areas of modern phenomenology where charm decays play an important 
role is the indirect search for physics beyond the Standard Model. Indeed, large 
statistics usually available in charm physics experiment makes it possible to
probe small effects that might be generated by the presence of
new physics particles and interactions. A program of searches for new physics 
in charm is complimentary to the corresponding programs in bottom or strange 
systems. This is in part due to the fact loop-dominated processes such as
$\DDbar$ mixing or flavor-changing neutral current (FCNC) decays are
sensitive to the dynamics of ultra-heavy {\it down-type particles}. Also, 
in many dynamical models, including the Standard Model, the effects in $s$, $c$, 
and $b$ systems are correlated.

The low energy effect of new physics particles can be 
naturally written in terms of a series of local operators of increasing
dimension generating $\Delta C = 1$ (decays) or $\Delta C = 2$ (mixing) 
transitions. For $\DDbar$ mixing these operators,
as well as the one loop Standard Model effects, generate contributions 
to the effective operators that change $D^0$ state into $\barD$ state
leading to the mass eigenstates
\begin{equation} \label{definition1}
| D_{^1_2} \rangle =
p | D^0 \rangle \pm q | \bar D^0 \rangle,
\end{equation}
where the complex parameters $p$ and $q$ are obtained from diagonalizing 
the $D^0-\barD$ mass matrix. The mass and width splittings between these 
eigenstates are given by
\begin{eqnarray} \label{definition}
x \equiv \frac{m_2-m_1}{\Gamma}, ~~
y \equiv \frac{\Gamma_2 - \Gamma_1}{2 \Gamma}.
\end{eqnarray}
It is known experimentally 
that $\DDbar$ mixing proceeds extremely slowly, which in the Standard Model 
is usually attributed to the absence of superheavy quarks destroying GIM 
cancellations~\cite{Petrov:1997ch}.

It is instructive to see how new physics can affect charm mixing.
Since the lifetime difference $y$ is constructed from the decays of $D$ into 
physical states, it should be dominated by the Standard Model contributions, 
unless new physics significantly modifies $\Delta C=1$ interactions. On the 
contrary, the mass difference $x$ can receive contributions from all energy scales.
Thus, it is usually conjectured that new physics can significantly
modify $x$ leading to the inequality $x\gg y$~\footnote{This signal for new physics 
is lost if a relatively large $y$, of the order of a percent, 
is observed~\cite{Bergmann:2000id,Falk:2001hx}.}. 
The same considerations apply to FCNC decays as well, where new physics could possibly 
contribute to the decay rates of $D \to X_u \gamma,~D \to X_u l^+ l^-$ 
(with $X_u$ being exclusive or inclusive final state) as well as other 
observables~\cite{FajferProc}. One technical problem here is that
in the standard model these decays are overwhelmingly dominated by long-distance
effects, which makes them extremely difficult to predict model-independently. 
This problem can be turned into a virtue~\cite{Fajfer:2000zx}.

Another possible manifestation of new physics interactions in the charm
system is associated with the observation of (large) CP-violation. This 
is due to the fact that all quarks that build up the hadronic states in weak 
decays of charm mesons belong to the first two generations. Since $2\times2$ 
Cabbibo quark mixing matrix is real, no CP-violation is possible in the
dominant tree-level diagrams that describe the decay amplitudes. 
In the Standard Model CP-violating amplitudes can be introduced by including 
penguin or box operators induced by virtual $b$-quarks. However, their 
contributions are strongly suppressed by the small combination of 
CKM matrix elements $V_{cb}V^*_{ub}$. It is thus widely believed that the 
observation of (large) CP violation in charm decays or mixing would be an 
unambiguous sign for new physics. This fact makes charm decays a valuable 
tool in searching for new physics, since the statistics available in charm 
physics experiment is usually quite large.

As in B-physics, CP-violating contributions in charm can be generally 
classified by three different categories:
(I) CP violation in the decay amplitudes. This type of CP violation 
occurs when the absolute value of the decay amplitude for $D$ to decay to a 
final state $f$ ($A_f$) is different from the one of corresponding 
CP-conjugated 
amplitude (``direct CP-violation'');
(II) CP violation in $\DDbar$ mixing matrix. This type of CP violation
is manifest when 
$R_m^2=\left|p/q\right|^2=(2 M_{12}-i \Gamma_{12})/(2 M_{12}^*-i 
\Gamma_{12}^*) \neq 1$; 
and 
(III) CP violation in the interference of decays with and without mixing.
This type of CP violation is possible for a subset of final states to which
both $D^0$ and $\barD$ can decay. 

For a given final state $f$, CP violating contributions can be summarized 
in the parameter 
\begin{equation}
\lambda_f = \frac{q}{p} \frac{{\overline A}_f}{A_f}=
R_m e^{i(\phi+\delta)}\left| \frac{{\overline A}_f}{A_f}\right|,
\end{equation}
where $A_f$ and ${\overline A}_f$ are the amplitudes for $D^0 \to f$ and 
$\barD \to f$ transitions respectively and $\delta$ is the strong phase 
difference between $A_f$ and ${\overline A}_f$. Here $\phi$ represents the
convention-independent weak phase difference between the ratio of 
decay amplitudes and the mixing matrix.

Presently, experimental information about the $\DDbar$ mixing parameters 
$x$ and $y$ comes from the time-dependent analyses that can roughly be divided
into two categories. First, more traditional studies look at the time
dependence of $D \to f$ decays, where $f$ is the final state that can be
used to tag the flavor of the decayed meson. The most popular is the
non-leptonic doubly Cabibbo suppressed decay $D^0 \to K^+ \pi^-$.
Time-dependent studies allow one to separate the DCSD from the mixing 
contribution $D^0 \to \D0bar \to K^+ \pi^-$,
\begin{eqnarray}\label{Kpi}
\Gamma[D^0 \to K^+ \pi^-]
=e^{-\Gamma t}|A_{K^-\pi^+}|^2 
~\left[
R+\sqrt{R}R_m(y'\cos\phi-x'\sin\phi)\Gamma t
+\frac{R_m^2}{4}(y^2+x^2)(\Gamma t)^2
\right],
\end{eqnarray}
where $R$ is the ratio of DCS and Cabibbo favored (CF) decay rates. 
Since $x$ and $y$ are small, the best constraint comes from the linear terms 
in $t$ that are also {\it linear} in $x$ and $y$.
A direct extraction of $x$ and $y$ from Eq.~(\ref{Kpi}) is not possible due 
to unknown relative strong phase $\delta_D$ of DCS and CF 
amplitudes~\cite{Falk:1999ts}, 
as $x'=x\cos\delta_D+y\sin\delta_D$, $y'=y\cos\delta_D-x\sin\delta_D$. 
This phase can be measured independently. The corresponding formula can 
also be written~\cite{Bergmann:2000id} for $\barD$ decay with $x' \to -x'$ and 
$R_m \to R_m^{-1}$.

Second, $D^0$ mixing can be measured by comparing the lifetimes 
extracted from the analysis of $D$ decays into the CP-even and CP-odd 
final states. This study is also sensitive to a {\it linear} function of 
$y$ via
\begin{equation}
\frac{\tau(D \to K^-\pi^+)}{\tau(D \to K^+K^-)}-1=
y \cos \phi - x \sin \phi \left[\frac{R_m^2-1}{2}\right].
\end{equation}
Time-integrated studies of the semileptonic transitions are sensitive
to the {\it quadratic} form $x^2+y^2$ and at the moment are not 
competitive with the analyses discussed above. 

The construction of new tau-charm factories CLEO-c and 
BES-III will introduce new {\it time-independent} methods that 
are sensitive to a linear function of $y$. One can again use the 
fact that heavy meson pairs produced in the decays of heavy quarkonium 
resonances have the useful property that the two mesons are in the CP-correlated 
states~\cite{AtwoodPetrov}.

By tagging one of the mesons as a CP eigenstate, a lifetime difference 
may be determined by measuring the leptonic branching ratio of the other meson.
Its semileptonic {\it width} should be independent of the CP quantum number 
since it is flavor specific, yet its {\it branching ratio} will be inversely 
proportional to the total width of that meson. Since we know whether this $D(k_2)$ state is 
tagged as a (CP-eigenstate) $D_\pm$ from the decay of $D(k_1)$ to a 
final state $S_\sigma$ of definite CP-parity $\sigma=\pm$, we can 
easily determine $y$ in terms of the semileptonic branching ratios of $D_\pm$. This 
can be expressed simply by introducing the ratio
\begin{equation} \label{DefCor}
R^L_\sigma=
\frac{\Gamma[\psi_L \to (H \to S_\sigma)(H \to X l^\pm \nu )]}{
\Gamma[\psi_L \to (H \to S_\sigma)(H \to X)]~Br(H^0 \to X l \nu)},
\end{equation}
where $X$ in $H \to X$ stands for an inclusive set of all
final states. A deviation from $R^L_\sigma=1$ implies a
lifetime difference. Keeping only the leading (linear) contributions
due to mixing, $y$ can be extracted from this experimentally obtained 
quantity,
\begin{eqnarray}
y\cos\phi=
(-1)^L {\sigma}
{R^L_\sigma-1\over R^L_\sigma}
\label{y-cos-phi}.
\end{eqnarray}

The current experimental upper bounds on $x$ and $y$ are on the order of 
a few times $10^{-2}$, and are expected to improve significantly in the coming
years.  To regard a future discovery of nonzero $x$ or $y$ as a signal for new 
physics, we would need high confidence that the Standard Model predictions lie
well below the present limits.  As was recently shown~\cite{Falk:2001hx}, 
in the Standard Model, $x$ and $y$ are generated only at second order in SU(3)$_F$ 
breaking, 
\begin{equation}
x\,,\, y \sim \sin^2\theta_C \times [SU(3) \mbox{ breaking}]^2\,,
\end{equation}
where $\theta_C$ is the Cabibbo angle.  Therefore, predicting the
Standard Model values of $x$ and $y$ depends crucially on estimating the 
size of SU(3)$_F$ breaking.  Although $y$ is expected to be determined
by the Standard Model processes, its value nevertheless affects significantly 
the sensitivity to new physics of experimental analyses of $D$ 
mixing~\cite{Bergmann:2000id}.

\begin{figure}[htb]
\begin{center}
\epsfig{file=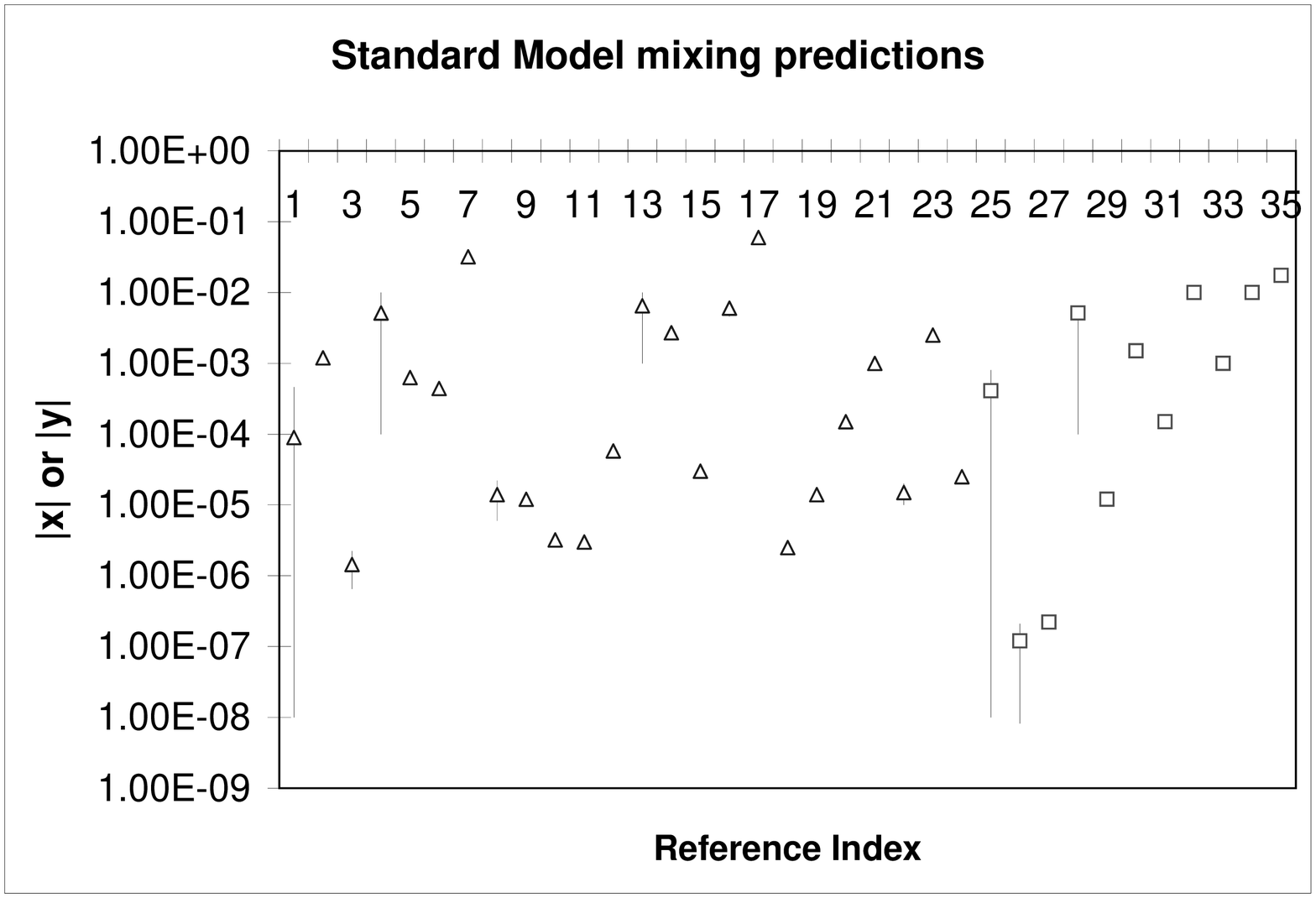,height=10cm,width=12cm}
\caption{Standard Model predictions for $|x|$ (open triangles) 
and $|y|$ (open squares). Horizontal line references are tabulated in Table~\ref{tab4}.}
\label{fig1}
\end{center}
\end{figure}

\begin{figure}[htb]
\begin{center}
\epsfig{file=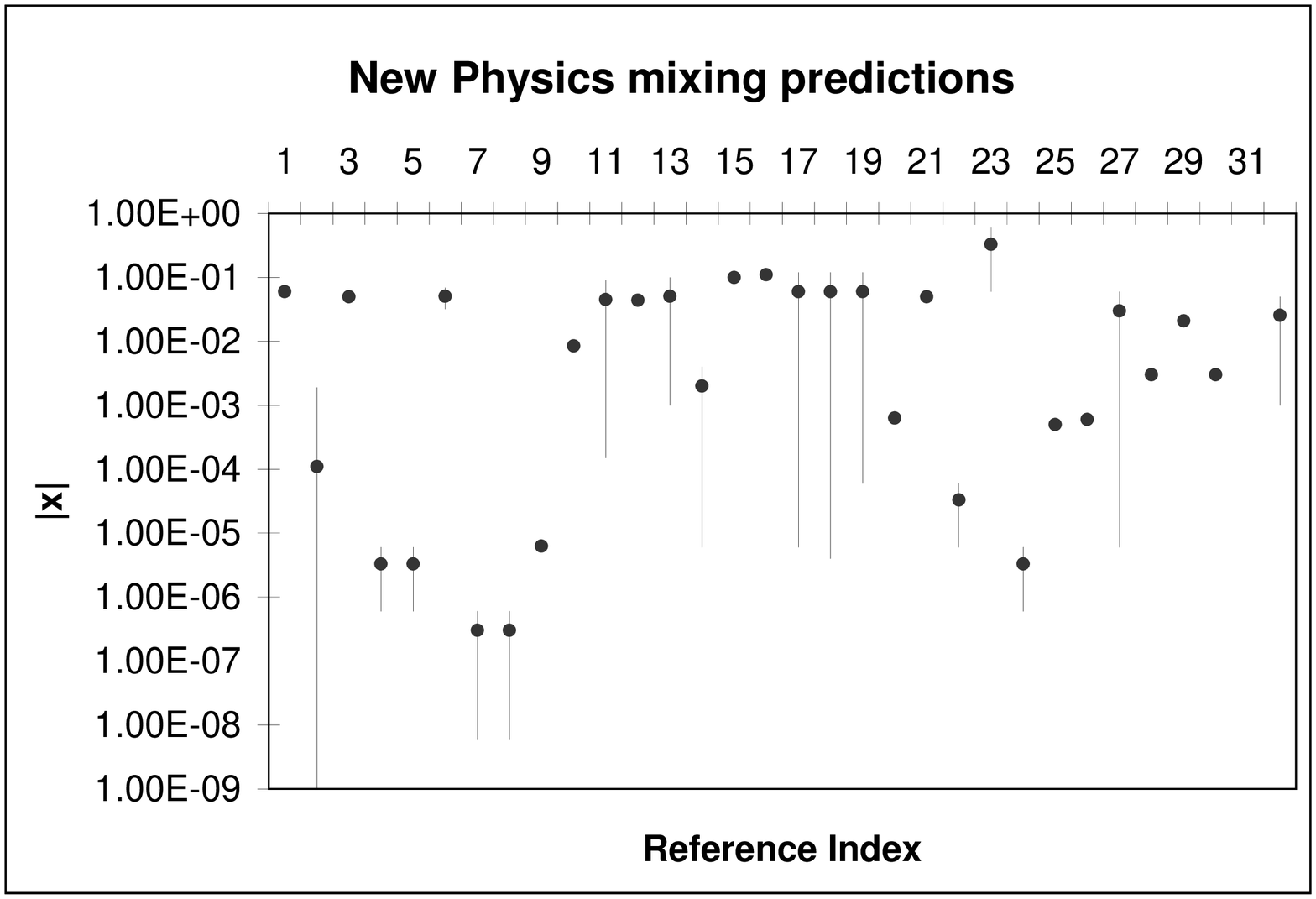,height=10cm,width=12cm}
\caption{New Physics predictions for $|x|$. Horizontal line references are 
tabulated in Table~\ref{tab5}.}
\label{fig2}
\end{center}
\end{figure}

Theoretical predictions of $x$ and $y$ span several orders of magnitude.
The predictions obtained in the framework of the Standard Model are not exception, 
as evidenced from Fig.~\ref{fig1}\footnote{Compilation of the $\DDbar$ mixng predictions 
is known as the Nelson plot~\cite{Nelson:1999fg}. In order to
obtain a compilation of the Standard Model (Fig.~\ref{fig1})
and new physics (Fig.~\ref{fig2}) predictions for charm mixing,
we updated and corrected Ref.~\cite{Nelson:1999fg} to remove double counting 
of predictions. We also separated the Standard Model and new physics predictions into
two separate plots~\cite{Improved}.}. Roughly, there are two approaches, neither of which 
give very reliable results because $m_c$ is in some sense intermediate between 
heavy and light.  The ``inclusive'' approach is based on the operator
product expansion (OPE).  In the $m_c \gg \Lambda$ limit, where
$\Lambda$ is a scale characteristic of the strong interactions, $\Delta
M$ and $\Delta\Gamma$ can be expanded in terms of matrix elements of local
operators~\cite{Inclusive}.  Such calculations yield $x,y < 10^{-3}$.  
The use of the OPE relies on local quark-hadron duality, 
and on $\Lambda/m_c$ being small enough to allow a truncation of the series
after the first few terms.  The charm mass may not be large enough for these 
to be good approximations, especially for nonleptonic $D$ decays.
An observation of $y$ of order $10^{-2}$ could be ascribed to a
breakdown of the OPE or of duality,  but such a large
value of $y$ is certainly not a generic prediction of OPE analyses.
The ``exclusive'' approach sums over intermediate hadronic
states, which may be modeled or fit to experimental data~\cite{Exclusive}.
Since there are cancellations between states within a given $SU(3)$
multiplet, one needs to know the contribution of each state with high 
precision. However, the $D$ is not light enough that its decays are dominated
by a few final states.  In the absence of sufficiently precise data on many decay 
rates and on strong phases, one is forced to use some assumptions. While most 
studies find $x,y < 10^{-3}$, Refs.~\cite{Exclusive} obtain $x$ and 
$y$ at the $10^{-2}$ level by arguing that SU(3)$_F$ violation is of order
unity, but the source of the large SU(3)$_F$ breaking is not made explicit.
It was also shown that phase space effects alone provide enough SU(3)$_F$ 
violation to induce $y\sim10^{-2}$~\cite{Falk:2001hx}.
Large effects in $y$ appear for decays close to $D$ threshold, where
an analytic expansion in SU(3)$_F$ violation is no longer possible.
Thus, theoretical calculations of $x$ and $y$ are quite uncertain, and the values
near the current experimental bounds cannot be ruled out. Therefore, it will 
be difficult to find a clear indication of 
physics beyond the Standard Model in $\DDbar$ mixing measurements alone.
The only robust potential signal of new physics in charm system at this stage 
is CP violation.

\section*{Acknowledgments}
This research was supported by the National Science Foundation under 
Grant PHY-0244853 and by the US Department of Energy under grant 
DE-FG02-96ER41005.


%
\begin{table}[htbp]
\begin{center}
\begin{tabular}{|c|c|c|}
\hline
Mass difference& Refrence & Citation \\
$x$ & Index & $~$ \\
\hline
\hline
$(0.9\pm 3.7)\times 10^{-4}$ & 1 & Phys. Rev. D 26, 143 (1982) \\
$1.2\times 10^{-3}$ & 2 & Phys. Lett. B128, 240 (1983) \\
$(1.44\pm 0.79)\times 10^{-6}$ & 3 & Z. Phys. C 27, 515 (1985) \\
$(0.01-10)\times 10^{-2}$ & 4 & Phys. Lett. B 164, 170 (1985) \\
$6.3\times 10^{-4}$ & 5 & Phys. Rev. D 33, 179 (1986) \\
$4.4\times 10^{-4}$ & 6 & Phys. Rev. D 35, 3484 (1987) \\
$3.2\times 10^{-2}$ & 7 & Phys. Lett. B224, 71 (1990) \\
$(1.4\pm 0.8)\times 10^{-5}$ & 8 & Nucl. Phys. B403, 71 (1993) \\
$1.2\times 10^{-5}$ & 9 & hep-ph/9407378\\
$3.2\times 10^{-6}$ & 10 & Chin. J. Phys. 32, 1163 (1994)  \\
$3.0\times 10^{-6}$ & 11 & hep-ph/9409379 \\
$5.8\times 10^{-5}$ & 12 & hep-ph/9508349\\
$(1-10)\times 10^{-3}$ & 13 & hep-ph/9508349\\
$2.7\times 10^{-4}$ & 14 & hep-ph/9508349\\
$3\times 10^{-5}$ & 15 & Phys. Lett. B357, 151 (1995) \\
$(6.0\pm 1.4)\times 10^{-3}$ & 16 & Phys. Lett. B357, 151 (1995) \\
$6\times 10^{-2}$ & 17 & Phys. Lett. B357, 151 (1995) \\
$2.5\times 10^{-6}$ & 18 & Phys. Rev. D 56, 1685 (1997) \\
$1.4\times 10^{-5}$ & 19 & Phys. Lett. B 422, 265 (1998) \\
$1.5\times 10^{-4}$ & 20 & Phys. Lett. B 427, 172 (1998) \\
$1.0\times 10^{-3}$ & 21 & Nucl.Phys. B592, 92 (2001) \\
$(1.5\pm 0.5)\times 10^{-5}$ & 22 & Phys. Lett. B297, 353 (1992) \\
$2.50\times 10^{-3}$ & 23 & Phys. Rev. D 43, 1641 (1991) \\
$2.50\times 10^{-5}$ & 24 & hep-ph/9706548 \\
\hline
\hline
Lifetime difference & Refrence & Citation \\
$y$ & Index & $~$ \\
\hline
\hline
$-(0.06-8.0)\times 10^{-4}$ & 25 & Phys. Rev. D 26, 143 (1982) \\
$(0.082-2.1)\times 10^{-7}$ & 26 & Phys. Lett. B 128, 240 (1983) \\
$2.2\times 10^{-7}$ & 27 & Z. Phys. C 27, 515 (1985) \\
$(0.01-10)\times 10^{-2}$ & 28 & Phys. Lett. B 164, 170 (1985) \\
$1.2\times 10^{-5}$ & 29 & hep-ph/9407378 \\
$1.5\times 10^{-3}$ & 30 & Phys. Lett. B 379, 249 (1996) \\
$1.0\times 10^{-4}$ & 31 & Phys. Lett. B 427, 172 (1998) \\
$1.0\times 10^{-2}$ & 32 & Phys. Rev. Lett. 83, 4005 (1999) \\ 
$1.0\times 10^{-3}$ & 33 & Nucl.Phys. B592, 92 (2001) \\
$1.0\times 10^{-2}$ & 34 & Phys. Rev. D65, 054034 (2002) \\ 
$(1.5-2.0)\times 10^{-2}$ & 35 & Phys. Rev. D 43, 1641 (1991) \\
\hline
\end{tabular}
\caption{Theoretical predictions for mixing parameters (Standard Model). The notation 
``$\pm$'' indicates the range of predictions.} 
\label{tab4}
\end{center}
\end{table}
\begin{table}[htbp]
\begin{center}
\begin{tabular}{|c|c|c|}
\hline
Mass difference& Refrence & Citation \\
$x$ & Index & $~$ \\
\hline
\hline
$6.0\times 10^{-2}$ & 1 & Yad. Phys. 34, 435 (1981) \\
$(0.11 \pm 1.8)\times 10^{-3}$ & 2 & Phys. Rev. D 26, 143 (1982) \\
$5\times 10^{-2}$ & 3 & Phys. Lett. B154, 287 (1985)  \\
$(0.6-6.0)\times 10^{-5}$ & 4 & Phys. Lett. B154, 287 (1985) \\
$(0.6-6.0)\times 10^{-4}$ & 5 & Phys. Lett. B154, 287 (1985) \\
$(5.05\pm 1.85)\times 10^{-2}$ & 6 & Phys. Lett. B154, 287 (1985) \\
$(0.06-60)\times 10^{-8}$ & 7 & Phys. Lett. B154, 287 (1985) \\
$(0.06-60)\times 10^{-5}$ & 8 & Phys. Lett. B154, 287 (1985) \\
$6.3\times 10^{-6}$ & 9 & Z. Phys. C 30, 293 (1986) \\
$8.5\times 10^{-3}$ & 10 & Z. Phys. C 30, 293 (1986) \\
$(0.15-90)\times 10^{-3}$ & 11 & Phys. Lett. B190, 93 (1987) \\
$4.4\times 10^{-2}$ & 12 & Phys. Rev. D 35, 3484 (1987)\\
$(0.1-10)\times 10^{-2}$ & 13 & Phys. Lett. B205, 540 (1988)\\
$(0.06-40)\times 10^{-4}$ & 14 & Phys. Rev. D 39, 878 (1989)\\
$0.1$ & 15 & Phys. Lett. B309, 337 (1993) \\
$0.11$ & 16 & Phys. Rev. D 48, 979 (1993) \\
$(0.006-120)\times 10^{-3}$ & 17 & hep-ph/9409379 \\
$(0.004-120)\times 10^{-3}$ & 18 & hep-ph/9409379 \\
$(0.06-120)\times 10^{-3}$ & 19 & hep-ph/9409379 \\
$6.3\times 10^{-4}$ & 20 & hep-ph/9409379 \\
$5\times 10^{-2}$ & 21 & hep-ph/9508349 \\
$(0.6-6)\times 10^{-5}$ & 22 & Phys. Rev. D 55, 3156 (1997) \\
$(0.6-6)\times 10^{-1}$ & 23 & Phys. Rev. D 55, 3156 (1997) \\
$(0.6-6)\times 10^{-6}$ & 24 & Phys. Rev. D 55, 3156 (1997) \\
$5\times 10^{-4}$ & 25 & Phys. Rev. Lett. 78, 2300 (1997) \\
$6\times 10^{-4}$ & 26 & hep-ph/9704316 \\
$(0.06-600)\times 10^{-4}$ & 27 & Phys. Lett. B416, 184 (1998) \\
$3\times 10^{-3}$ & 28 & Phys. Rev. D 60, 013005 (1999) \\
$2.1\times 10^{-2}$ & 29 & Phys. Rev. D 61, 075011 (2000) \\
$3.0\times 10^{-2}$ & 30 & Phys. Rev. D 62, 033002 (2000) \\
$0.001-0.05$ & 31 & hep-ph/0110106 \\
\hline
\end{tabular}
\caption{Theoretical predictions for mixing parameters (New Physics). The notation 
``$\pm$'' indicates the range of predictions based on the model parameter space bounded by 
the data available {\it at the time of publication}.} 
\label{tab5}
\end{center}
\end{table}
%


%
%


%
\label{PetrovEnd}
 

\begin{thebibliography}{99}


\bibitem{Petrov:2003rm}
A.~A.~Petrov,
eConf {\bf C0304052}, WG506 (2003)
[arXiv:hep-ph/0307322].

\bibitem{chistov-fpcp03}  R.~Chistov, ``Charm Spectroscopy'', {\it these proceedings}.

\bibitem{johns-fpcp03}  W.~Johns, ``Semileptonic and rare charm decays'', 
{\it these proceedings.}

\bibitem{boca-fpcp03}  G.~Boca, ``$D$ mixing and lifetimes'', {\it these proceedings.}

\bibitem{dspi0_babar} B. Aubert {\it et al.} (BaBar Collaboration),
Phys. Rev. Lett. {\bf 90}, 242001 (2003).

\bibitem{dspi0_cleo} D. Besson {\it et al.} 
(CLEO Collaboration), hep-ex/0305100, submitted to Phys. Rev. D.

\bibitem{dspi0_belle}
K. Abe {\it et al.} (Belle Collaboration), BELLE-CONF-0340,
BELLE-CONF-0334, contributed to the EPS and LP03 conferences,
http://belle.kek.jp/conferences/LP03-EPS.


\bibitem{PDG} D.E. Groom {\it et al.} (Particle Data Group), 
Eur. Phys. J. C {\bf 15}, 1 (2000).

\bibitem{godfrey_isgur} 
S.~Godfrey and N.~Isgur,
Phys.\ Rev.\ D {\bf 32}, 189 (1985).

\bibitem{eichten} 
M.~Di Pierro and E.~Eichten,
Phys.\ Rev.\ D {\bf 64}, 114004 (2001).

\bibitem{Gupta:1994mw}
S.~N.~Gupta and J.~M.~Johnson,
Phys.\ Rev.\ D {\bf 51}, 168 (1995).

\bibitem{Zeng:1994vj}
J.~Zeng, J.~W.~Van Orden and W.~Roberts,
Phys.\ Rev.\ D {\bf 52}, 5229 (1995).

\bibitem{Kalashnikova:2001ig}
Y.~S.~Kalashnikova, A.~V.~Nefediev and Y.~A.~Simonov,
Phys.\ Rev.\ D {\bf 64}, 014037 (2001).

\bibitem{Ebert:1997nk}
D.~Ebert, V.~O.~Galkin and R.~N.~Faustov,
Phys.\ Rev.\ D {\bf 57}, 5663 (1998)
[Erratum-ibid.\ D {\bf 59}, 019902 (1999)].

\bibitem{Lewis:2000sv}
R.~Lewis and R.~M.~Woloshyn,
Phys.\ Rev.\ D {\bf 62}, 114507 (2000);
J.~Hein {\it et al.},
Phys.\ Rev.\ D {\bf 62}, 074503 (2000);
G.~S.~Bali, Phys.\ Rev.\ D {\bf 68}, 071501 (2003)
[arXiv:hep-ph/0305209].

\bibitem{lattice} A.~Dougall, R.~D.~Kenway, C.~M.~Maynard and C.~McNeile,
Phys.\ Lett.\ B {\bf 569}, 41 (2003)
[arXiv:hep-lat/0307001].

\bibitem{chiral}
W.~A.~Bardeen and C.~T.~Hill,
Phys.\ Rev.\ D {\bf 49}, 409 (1994);
W.~A.~Bardeen, E.~J.~Eichten and C.~T.~Hill,
Phys.\ Rev.\ D {\bf 68}, 054024 (2003)
[arXiv:hep-ph/0305049];
D.~Ebert, T.~Feldmann and H.~Reinhardt,
Phys.\ Lett.\ B {\bf 388}, 154 (1996);
D.~Ebert, T.~Feldmann, R.~Friedrich and H.~Reinhardt,
Nucl.\ Phys.\ B {\bf 434}, 619 (1995);
M.~A.~Nowak, M.~Rho and I.~Zahed,
Phys.\ Rev.\ D {\bf 48}, 4370 (1993);
M.~A.~Nowak, M.~Rho and I.~Zahed,
arXiv:hep-ph/0307102;
A.~Deandrea, G.~Nardulli and A.~D.~Polosa,
Phys.\ Rev.\ D {\bf 68}, 097501 (2003)
[arXiv:hep-ph/0307069].

\bibitem{barnes_lipkin} T.~Barnes, F.~E.~Close and H.~J.~Lipkin, 
Phys.\ Rev.\ D {\bf 68}, 054006 (2003)
[arXiv:hep-ph/0305025].

\bibitem{cheng_hou} H.~Y.~Cheng and W.-S.~Hou, Phys.\ Lett.\ B {\bf 566}, 193 (2003)
[arXiv:hep-ph/0305038].

\bibitem{szcz} A.~P.~Szczepaniak, Phys.\ Lett.\ B {\bf 567}, 23 (2003)
[arXiv:hep-ph/0305060].

\bibitem{cahn_jackson}
R.~N.~Cahn and J.~D.~Jackson,
Phys.\ Rev.\ D {\bf 68}, 037502 (2003)
[arXiv:hep-ph/0305012];
Y.~B.~Dai, C.~S.~Huang, C.~Liu and S.~L.~Zhu,
arXiv:hep-ph/0306274.

\bibitem{ddst_belle}
K. Abe {\it et al.} (Belle Collaboration), hep-ex/0307021, submitted
to Phys. Rev. D.

\bibitem{beveren} E. van Beveren et al., Z. Phys. C 30, 615 (1986);
E. van Beveren and G. Rupp, hep-ph/0304105; E. van Beveren and
G. Rupp, Phys. Rev. Lett. 91, 012003 (2003);
J. Weinstein and N. Isgur, Phys. Rev. D 41, 2236 (1990).

\bibitem{lipkin_1} H.J. Lipkin, Phys. Lett. B. 70, 113 (1977).

\bibitem{suzuki} M. Suzuki and S.F. Tuan, Phys. Lett. B. 133, 125 (1983).

\bibitem{Godfrey:2003kg}
S.~Godfrey,
Phys.\ Lett.\ B {\bf 568}, 254 (2003)
[arXiv:hep-ph/0305122];
P.~Colangelo and F.~De Fazio,
Phys.\ Lett.\ B {\bf 570}, 180 (2003)
[arXiv:hep-ph/0305140].

\bibitem{Browder:2003fk}
T.~E.~Browder, S.~Pakvasa and A.~A.~Petrov,
arXiv:hep-ph/0307054, Phys. Lett. B, in press.

\bibitem{Lucha:2003gs}
W.~Lucha and F.~F.~Schoberl,
arXiv:hep-ph/0309341.

\bibitem{Datta:2003re}
A.~Datta and P.~J.~O'donnell,
Phys.\ Lett.\ B {\bf 572}, 164 (2003)
[arXiv:hep-ph/0307106].

\bibitem{veseli} S.~Veseli and I.~Dunietz,
Phys.\ Rev.\ D {\bf 54}, 6803 (1996).



\bibitem{Buchalla:1995vs}
G.~Buchalla, A.~J.~Buras and M.~E.~Lautenbacher,
Rev.\ Mod.\ Phys.\  {\bf 68}, 1125 (1996);
A.~J.~Buras, M.~Jamin and P.~H.~Weisz,
Nucl.\ Phys.\ B {\bf 347}, 491 (1990);
M.~Ciuchini, E.~Franco, G.~Martinelli, L.~Reina and L.~Silvestrini,
Z.\ Phys.\ C {\bf 68}, 239 (1995).


\bibitem{Battaglia:2003in}
M.~Battaglia {\it et al.},
arXiv:hep-ph/0304132.

\bibitem{Aoki:2003xb}
S.~Aoki {\it et al.}  [JLQCD Collaboration],
arXiv:hep-ph/0307039.

\bibitem{Korner:2003zk}
J.~G.~Korner, A.~I.~Onishchenko, A.~A.~Petrov and A.~A.~Pivovarov,
Phys.\ Rev.\ Lett.\  {\bf 91}, 192002 (2003)
[arXiv:hep-ph/0306032].


\bibitem{Grinstein:1993ys}
B.~Grinstein,
Phys.\ Rev.\ Lett.\  {\bf 71}, 3067 (1993).

\bibitem{Ligeti:1997aq}
Z.~Ligeti, I.~W.~Stewart and M.~B.~Wise,
Phys.\ Lett.\ B {\bf 420}, 359 (1998).

\bibitem{Pedrini:2003ee}
D.~Pedrini  [the FOCUS Collaboration],
arXiv:hep-ph/0307137;
M.~B.~Voloshin,
Phys.\ Rept.\  {\bf 320}, 275 (1999);
B.~Guberina, B.~Melic and H.~Stefancic,
Phys.\ Lett.\ B {\bf 484}, 43 (2000).

\bibitem{Juttner:2003cf}
A.~Juttner and J.~Rolf,
arXiv:hep-ph/0306299;




\bibitem{Petrov:1997ch}
A.~Datta, D.~Kumbhakar,
Z.\ Phys.\ C{\bf 27}, 515 (1985);
A.~A.~Petrov,
Phys.\ Rev.\ D{\bf 56}, 1685 (1997);
E.~Golowich and A.~A.~Petrov,
Phys.\ Lett.\ B {\bf 427}, 172 (1998).

\bibitem{Bergmann:2000id}
S.~Bergmann, Y.~Grossman, Z.~Ligeti, Y.~Nir, A.~Petrov,
Phys.\ Lett.\ B {\bf 486}, 418 (2000).

\bibitem{Falk:2001hx}
A.~F.~Falk, Y.~Grossman, Z.~Ligeti and A.~A.~Petrov,
Phys.\ Rev.\ D {\bf 65}, 054034 (2002).

\bibitem{FajferProc}
S.~Fajfer,
arXiv:hep-ph/0306263.

\bibitem{Fajfer:2000zx}
S.~Fajfer, S.~Prelovsek, P.~Singer and D.~Wyler,
Phys.\ Lett.\ B {\bf 487}, 81 (2000).

\bibitem{Falk:1999ts}
A.~F.~Falk, Y.~Nir and A.~A.~Petrov,
JHEP {\bf 9912}, 019 (1999).

\bibitem{AtwoodPetrov}
D.~Atwood and A.~A.~Petrov,
arXiv:hep-ph/0207165.

\bibitem{Nelson:1999fg}
H.~N.~Nelson,
in {\it Proc. of the 19th Intl. Symp. on Photon and Lepton 
Interactions at High Energy LP99 } ed. J.A. Jaros and M.E. Peskin,
arXiv:hep-ex/9908021.

\bibitem{Improved}
For the most updated version, see 
http://www.physics.wayne.edu/~apetrov/mixing/.

\bibitem{Inclusive}
H.~Georgi, Phys. Lett. B297, 353 (1992);
T.~Ohl, G.~Ricciardi and E.~Simmons, Nucl. Phys. B403, 605 (1993);
I.~Bigi and N.~Uraltsev,
Nucl.\ Phys.\ B {\bf 592}, 92 (2001).

\bibitem{Exclusive}
J. Donoghue, E. Golowich, B. Holstein and J. Trampetic,
Phys. Rev. D33, 179 (1986);
L. Wolfenstein, Phys.\ Lett.\ B164, 170 (1985);
P. Colangelo, G. Nardulli and N. Paver,  Phys.\ Lett.\ B242, 71 (1990);
T.A. Kaeding,  Phys. Lett. B357, 151 (1995).
A.~A.~Anselm and Y.~I.~Azimov,
Phys.\ Lett.\ B {\bf 85}, 72 (1979);

\bibitem{BigiSandaBook}
I.~I.~Bigi and A.~I.~Sanda,
{\it CP violation} (Cambridge University Press, 2000).

\bibitem{Pedrini:2000ge}
D.~Pedrini,
J.\ Phys.\ G {\bf 27}, 1259 (2001).

\bibitem{Bigi:1986dp}
I.~I.~Bigi and A.~I.~Sanda,
Phys.\ Lett.\ B {\bf 171}, 320 (1986).

\bibitem{Petrov:2002qb}
A.~A.~Petrov,
Proc. of the {\it 5th Workshop on Continuous Advances in QCD}, 
pp. 102-114; arXiv:hep-ph/0209049.

\end{thebibliography}
\end{document}